# 3D Deep Learning Enables Fast Imaging of Spines through Scattering Media by Temporal Focusing Microscopy

Zhun Wei[1,†], Josiah R. Boivin[2,†], Yi Xue[3], Xudong Chen[4], Peter T. C. So[5,6,7], Elly Nedivi[2,8,9], Dushan N. Wadduwage[1,*]

[1]Center for Advanced Imaging, Faculty of Arts and Sciences, Harvard University, Cambridge, MA 02138, USA.

[2]Picower Institute for Learning and Memory, Massachusetts Institute of Technology, Cambridge, MA 02139, USA.

[3]Dept. of Electrical Engineering and Computer Sciences, University of California, Berkeley, CA 94720, USA.

[4]Department of Electrical and Computer Engineering, National University of Singapore, 4 Engineering Drive 3, Singapore 117583, Singapore.

[5]Dept. of Biological Engineering, [6]Laser Biomedical Research Center, [7]Dept. of Mechanical Engineering, Massachusetts Institute of Technology, 77 Massachusetts Ave., Cambridge, MA 02139, USA.

[8]Department of Brain and Cognitive Sciences, [9]Department of Biology, Massachusetts Institute of Technology, Cambridge, MA 02139, USA.

[†]These authors contributed equally to this work.

[*]Corresponding author: wadduwage@fas.harvard.edu

## Abstract

Today the gold standard for *in vivo* imaging through scattering tissue is the point-scanning two-photon microscope (PSTPM). Especially in neuroscience, PSTPM is widely used for deep-tissue imaging in the brain. However, due to sequential scanning, PSTPM is slow. Temporal focusing microscopy (TFM), on the other hand, focuses femtosecond pulsed laser light temporally, while keeping wide-field illumination, and is consequently much faster. However, due to the use of a camera detector, TFM suffers from the scattering of emission photons. As a result, TFM produces images of poor spatial resolution and signal-to-noise ratio (SNR), burying fluorescent signals from small structures such as dendritic spines. In this work, we present a data-driven deep learning approach to improve resolution and SNR of TFM images. Using a 3D convolutional neural network (CNN) we build a map from TFM to PSTPM modalities, to enable fast TFM imaging while maintaining high-resolution through scattering media. We demonstrate this approach for *in vivo* imaging of dendritic spines on pyramidal neurons in the mouse visual cortex. We show that our trained network rapidly outputs high-resolution images that recover biologically relevant features previously buried in the scattered fluorescence in the TFM images. *In vivo* imaging that combines TFM and the proposed 3D convolution neural network is one to two orders of magnitude faster than PSTPM but retains the high resolution and SNR necessary to analyze small fluorescent structures. The proposed 3D convolution deep network could also be potentially beneficial for improving the performance of many speed-demanding deep-tissue imaging applications such as *in vivo* voltage imaging.

**Introduction**

Imaging large 3D volumes at high spatial and temporal resolution is a fundamental challenge for *in vivo* microscopy. With the inherent advantages of high spatial resolution, low phototoxicity, and deep penetration, point-scanning two-photon microscopy (PSTPM) is currently the gold standard for *in vivo* imaging of fluorescently labelled structures[1-4]. However, due to sequential point scanning PSTPM is relatively slow, limiting the size, resolution, or the speed of volumetric imaging [5,6].

Temporal focusing microscopy (TFM)[7-9] overcomes the speed limitation of PSTPM by wide-field excitation and camera-based acquisition. Instead of focusing in space, TFM focuses an amplified femtosecond laser pulse in time. The pulse is first temporally dispersed using a diffraction grating, and then recreated solely at the focal plane, enabling depth selective two-photon excitation of millions of pixels simultaneously. While TFM excitation photons can penetrate through scattering tissue due to their long wavelengths, shorter-wavelengthed emission photons encounter significant scattering disturbances on their way to the detector; therefore, in a wide-field setting, some photons get mapped to incorrect detector pixels. Compared to PSTPM, TFM is thus much more sensitive to tissue scattering. The scattered photons degrade the spatial resolution and SNR, hindering TFM's ability to image small dim fluorescent structures. Thus, while TFM has been used successfully for *in vivo* imaging of large structures such as neuronal cell bodies at relatively low resolution[10,11], its use has not been demonstrated for *in vivo* imaging of small structures such as dendritic spines, whose relatively dim fluorescence can be lost to scattering. Overcoming the scattering limitation of TFM to achieve high-resolution imaging of large tissue volumes would open the door to a wealth of experiments that are infeasible with traditional PSTPM and TFM.

In most imaging modalities, such as TFM, the imaging process can be treated as a forward model ($f$) that transforms the ground truth image ($x$) to the observed image ($y$), where $y$ is usually degraded by noise contamination, scattering effects, low-pass filtering, and subsampling[12]. Reconstructing $x$ from the observed $y$ is a challenging, ill-posed inverse problem ($f^{-1}$) since the forward model could map multiple different images to the same observation image[13]. Traditionally, these ill-posed inverse problems are solved using slow iterative algorithms by considering *prior* information with regularizations[14]. However, the accuracy of these model-based algorithms is contingent upon successfully capturing the right prior information, a challenging task for complex image structures in many practical microscopy applications.

To incorporate *priors* for challenging inverse problems, machine learning has proven extremely capable, especially for vision tasks[12,15-17]. Recently, thanks to increased computational power and accumulated data, significant progress has been made in the field of machine learning, where a much deeper network can be used to achieve the state-of-the-art performance for many tasks. For example, deep learning has shown its success in image classification[18,19], prediction[20], segmentation[21,22], denoising[23], biomedical imaging[24-26], and other linear or nonlinear inverse problems[16,27-29]. In fluorescence microscopy, it is becoming an increasingly important tool for image reconstruction[12], image restoration[15], deconvolution[30], super-resolution[31-33], and other tasks that are either challenging or complicated using traditional techniques. It has been shown that, provided with sufficient training examples, deep learning is capable of extracting useful information[34] and recovering high-frequency information from raw data[35], which can be understood as a universal approximation[36].

De-scattering TFM images contaminated by tissue scattering is an interesting inverse problem. The extent of scattering gradually increases as the imaging depth is increased[3]. Little to no scattering is seen near the surface of the tissue; intuitively the surface plane can act as a strong prior to estimate the second *z*-plane. One could generalize this idea, to use a reconstructed *z*-plane as prior information to estimate the next *z*-plane from its TFM observation. More elegantly put, 3D features of image volumes can act as strong priors to de-scatter TFM images. Thus, in this work, we hypothesize that scattering is a gradually extending process (*f*), and learning 3D features can capture strong priors needed to approximate its inverse ($f^{-1}$). We first model the forward scattering process (*f*) with a scattering point-spread-function (sPSF) that gradually changes with *z*-depth[2]. We map a large library of domain specific PSTPM volumes through *f* to their corresponding TFM volumes. Then we train a deep 3D convolutional neural network to learn a volumetric inverse map, $f^{-1}$, from TFM to PSTPM images. Our 3D deep learning approach addresses the problem of low image quality in TFM. Rather than improving the performance using a mechanical approach, such as HiLL[37], mosTFM[38], TRAFIX[39] or DEEP-TFM[40] microscopy, we utilize a data-driven approach to recover meaningful patterns from high-speed TFM measurements. The results are verified by both numerical and *in vivo* experiments. Our results show that the deep learning approach can recover biologically meaningful features that were unrecognizable in the TFM images, including dendritic spines. By recovering small features that would otherwise be lost to scattering in TFM images, our deep learning approach enables rapid imaging of large 3D volumes at high resolution, nearly 50 times faster than PSTPM.

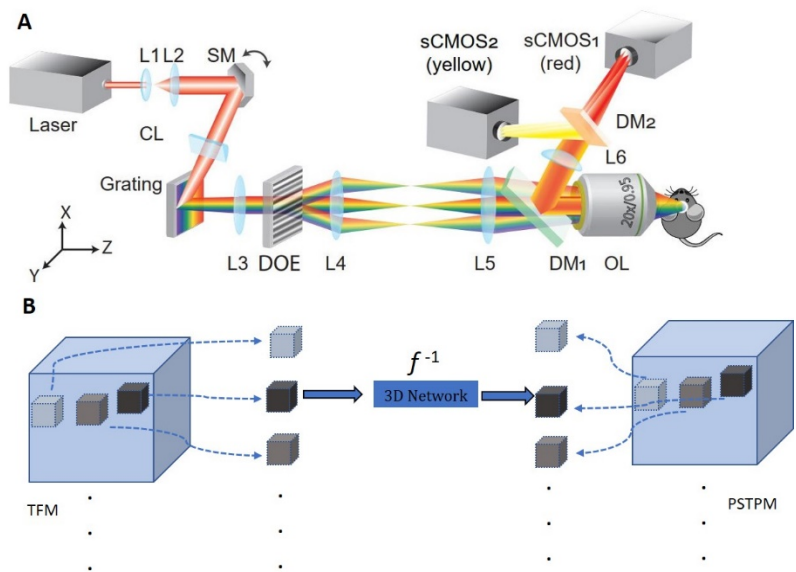

**Fig. 1** The optical schematic of the TFM setup (A): SM: scanning mirror. CL: cylindrical lens. Grating: dispersion along the x-axis. L3, L4: relay lenses. DOE: diffractive optical elements, to generate multiple lines for parallel scanning. L5, L6: tube lenses. DM: dichroic mirror. OL: objective lens. 3D deep learning model to solve the inverse problem $f^{-1}$, where 3D features of image volumes act as strong priors to de-scatter TFM images (B): cubes randomly generated from TFM image stacks with a 64 *μm* range in z-depth are used as input, and the 'ground truth' is the cubes from corresponding PSTPM images.

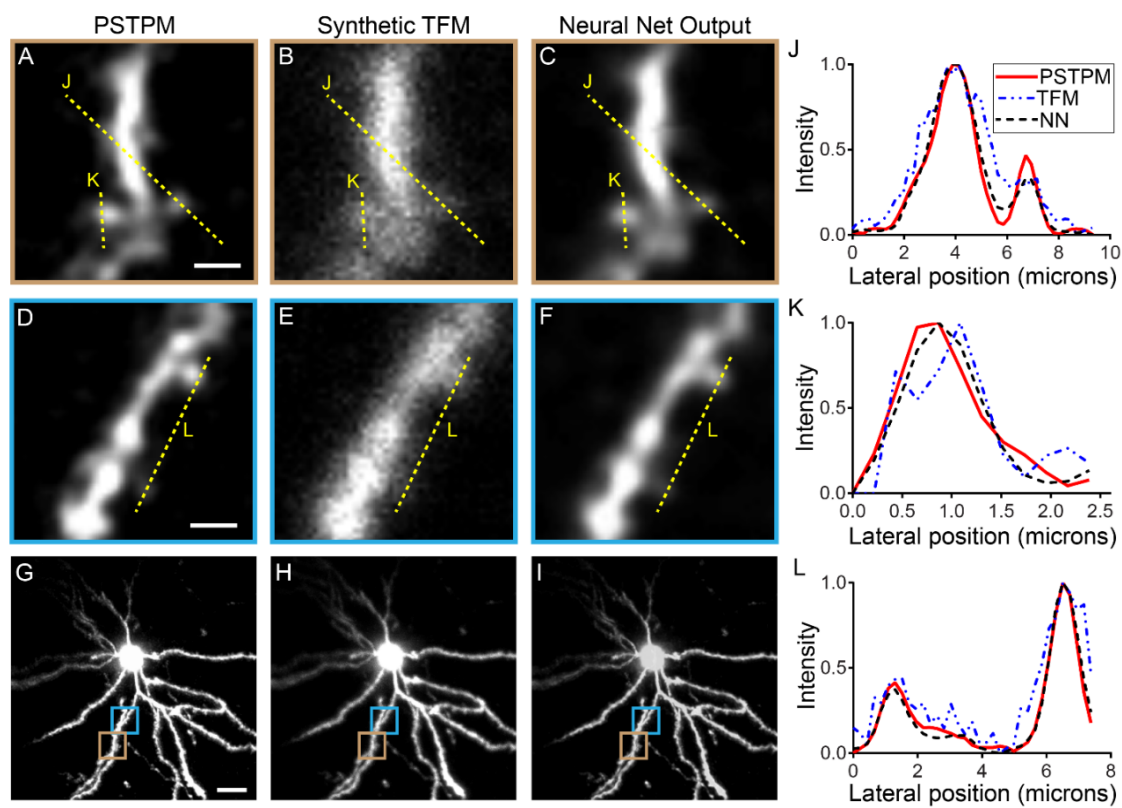

**Fig. 2** Numerical Results: Representative images of the dendrites of a neuron whose synthetic TFM image was used to test the performance of the neural network. Enlarged view of two example branches located at depths of 51 *μm* (A-C) and 34 *μm* (D-F) from the surface of the brain. The neuron's dendritic arbor was imaged *in vivo* using PSTPM (A, D, G). TFM images (B, E, H) were synthetically generated from PSTPM stacks, which are also the inputs of the neural network. The neural network was then able to extract relevant features from the synthetic TFM image (C, F, I), showing good correspondence with the 'ground truth' PSTPM image (A, D, G). Maximum intensity Z projection of the full 3D image stack is shown for the PSTPM (G), TFM (H), and neural net (I) images, with the example dendrites shown in A-F marked by color-coded boxes. Quantitative comparisons among PSTPM, TFM, and neural network output (neural net abbreviated as NN) are shown, where the normalized intensity profiles varying with the lateral positions are presented (J, K, L). The fluorescence intensity quantification shows clearly defined peaks at the sites of dendritic spines in the PSTPM and neural net output images, while TFM images either miss the peak (smaller peak in J) or show more peaks at incorrect positions (K, L). Scale bars: 2 *μm* for A-F, 20 *μm* for G-I.

## Results

**Numerical results:** To build a 3D deep learning model with PSTPM data, we synthetically generated TFM cubes from 33 PSTPM volumes that each included the dendritic arbor of a layer 2/3 pyramidal neuron imaged *in vivo* in the mouse brain (Methods). As presented in Fig. 1B, for each PSTPM volume, hundreds of smaller TFM cubes were synthetically generated for training purposes by randomly selecting the center position of these cubes.

Fig. 2 shows representative results of reconstructing dendrites of a neuron using the proposed deep learning model. In this example, the TFM inputs to the network were synthetically generated from a cell imaged with PSTPM; neither during training nor during validation had the network seen this cell. The proposed learning model reconstructs some important small features that have been either missing or hardly recognized in TFM results (Fig 2B,C and E, F). In Figs. 2J, K, and L, we also present the quantitative results by comparing the normalized intensity profiles in the lateral direction. The normalized intensity shows clearly defined peaks at the sites of dendritic spines in the PSTPM and neural net output images. However, in TFM images, these peaks are either obscured by scattered fluorescence along the dendritic shaft (in Fig. 2J) or distorted into multiple peaks (in Figs. 2K and L).

**Experimental results, dendritic spine analysis:** We next validated our method using experimental data. The apical dendritic tuft of a layer 2/3 pyramidal neuron in the mouse visual cortex was imaged *in vivo* using TFM. The same neuron was then imaged using PSTPM. TFM images served as the input for the trained network, while PSTPM images of the same dendrites served as 'ground truth' data to benchmark network performance. The TFM 3D image stack was divided into cubes and fed into the trained network. The reconstructed results generated by the trained network were then compared to the ground truth PSTPM image as shown in Fig. 3.

We quantified the ability of the trained network to reconstruct important features of the neuron by scoring dendritic spines across the full image stack. The image stack spanned a 200 by 200 *μm* field of view in XY and a 64 *μm* range in Z (depth: 14-78 *μm* from the surface of the brain), showing the fluorescently labelled apical dendritic tuft of the layer 2/3 pyramidal neuron with its resident spines. Dendritic spines were scored on the PSTPM image using established criteria based on a consensus of multiple laboratories that perform similar *in vivo* spine analysis[4,5]. Spines were defined as protrusions with a length of at least 0.75 *μm* (3 pixels in the PSTPM image) from the edge of the dendritic shaft, present in at least two consecutive Z planes. A total of 295 spines were scored in the PSTPM image. Each spine scored in the PSTPM image was then marked as present or absent in the TFM and neural net output images.

A representative dendritic branch at a depth of 66 *μm* from the surface of the brain is shown for the PSTPM image (Fig. 3A), TFM image (Fig. 3B), and neural net output image (Fig. 3C). Spines scored on the PSTPM image are denoted by green triangles in Fig. 3A. Each of these spines is then marked as present (filled green triangle) or absent (open red triangle) in the TFM and neural net output images (Fig. 3B-C). Two additional example branches are shown in Fig. 3D-F and Fig. 3G-I, with spines marked in the same manner. A maximum intensity Z projection of the full image stack is shown for the PSTPM image (Fig. 3J), TFM image (Fig. 3K), and neural net output image (Fig. 3L), with the location of each example branch indicated by color-coded rectangles. Of the 295 spines scored in the PSTPM image, 147 (49.8%) were visible in the TFM image, and 269 (91.2%) were visible in the neural net output image (Fig. 3M). (All spines that were visible on the TFM image were also visible on the neural net output image.) Our results indicate that the trained neural network can reconstruct dendritic spines that were previously lost in the scattered fluorescence of a TFM image, producing an image that shows close correspondence with the ground truth PSTPM image.

**Experimental results, fluorescence intensity quantification:** We further characterized our experimental results by quantifying the fluorescence intensity along regions of interest (ROIs) that included dendritic spines and the dendritic shaft (Fig. 4). In accordance with the spine analysis shown in Fig. 3, the fluorescence intensity quantification shows clearly defined peaks at the sites of dendritic spines in the PSTPM and neural net output images. These peaks are often obscured by surrounding noise from scattered fluorescence in the TFM images. For example, the PSTPM and neural net output traces in Fig. 4J show three peaks representing three dendritic spines, while the TFM trace shows only the brightest of these three spines; the dimmer two spines are obscured by scattered fluorescence along the dendritic shaft in the TFM trace (Fig. 4J).

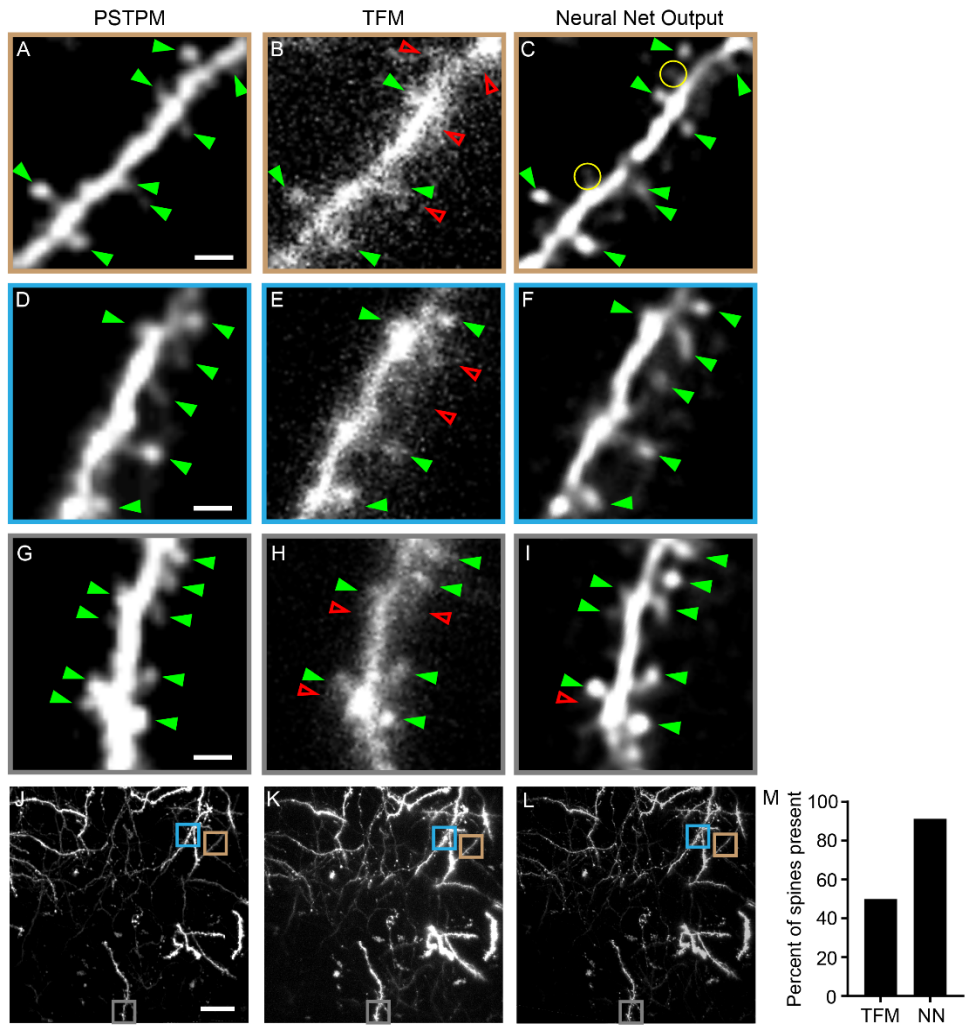

Fig. 3. Experimental Results: Representative images of the apical dendrites of a neuron whose TFM image was used to test neural network performance. Enlarged view of three example branches located at depths of 66 $\mu m$ (A-C), 41 $\mu m$ (D-F), and 64 $\mu m$ (G-I) from the surface of the brain. The neuron's apical dendritic tuft was imaged *in vivo* using PSTPM (A, D, G) and TFM (B, E, H). The neural network was then able to extract relevant features from the TFM image (C, F, I), showing good correspondence with the 'ground truth' PSTPM image (A, D, G). Green arrows denote dendritic spines scored on the PSTPM image (A, D, G). Each spine was then scored as present (green arrow) or absent (red open arrow) in the TFM (B, E, H) and neural net (C, F, I) images. Potential "false-positive" spines are marked in C by yellow circles and discussed further in Fig. 5. Maximum intensity Z projection of the full 3D image stack is shown for the PSTPM (J), TFM (K), and neural net (L) images, with the example dendrites shown in A-I marked by color-coded boxes. Of the 295 spines scored on the PSTPM image, 147 (49.8%) were visible in the TFM image, and 269 (91.2%) were visible in the neural net output image (M, neural net abbreviated as NN). Scale bars: 2 $\mu m$ for A-I, 20 $\mu m$ for J-L.

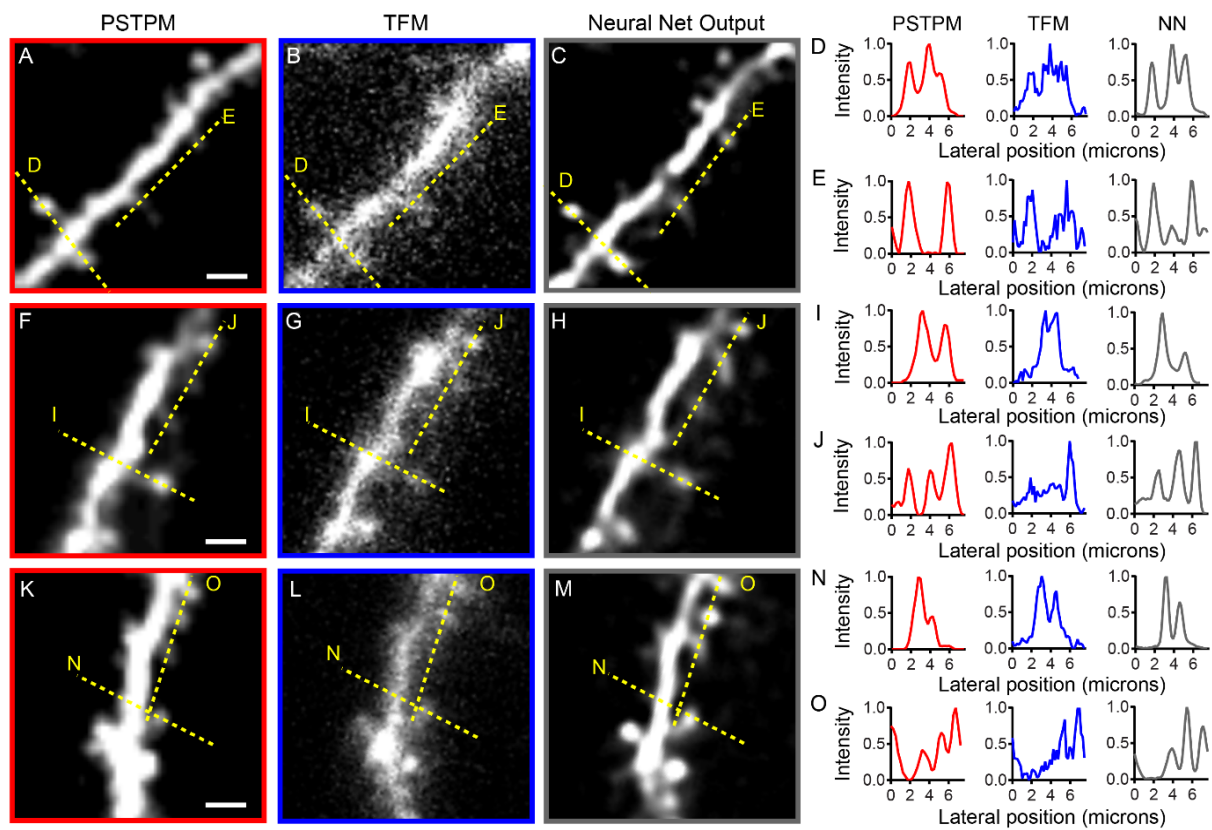

Fig. 4. Fluorescence intensity measurements for PSTPM, TFM and neural net (NN) output images. Images are the same as those shown in Fig. 3, with each row of images showing a segment of dendrite imaged using PSTPM (A, F, K) and TFM (B, G, L), then reconstructed from the TFM image using the artificial neural network (C, H, M). Fluorescence intensity was measured along the yellow dashed lines shown on the images, with traces of normalized intensity shown in the right panel. Scale bars, 2 $\mu m$.

**"False-Positive" analysis:** During the dendritic spine analysis, we found multiple dim spines present in the NN output image but not in the PSTPM image (e.g. Fig. 3C, yellow circles). To determine whether these spines represent structures missed in PSTPM imaging rather than artifacts generated by the trained network, we conducted the following experimental and numerical tests (also see detailed analysis in Supplementary Section 3).

Dim spines can be missed in PSTPM images when excitation power is relatively low, but would be revealed when excitation power is increased. The high excitation power in TFM could potentially excite fluorescent molecules in such small structures, although the signal would be obscured by the low SNR without application of the deep network analysis. To examine this possibility, dendritic segments were imaged on the PSTPM system using different excitation powers to determine whether dim spines might be missed using our usual imaging parameters. As shown for an example branch in Fig. 5, a dim spine is visible when imaged at 14 mW excitation, but not when imaged at 10 mW excitation (green triangle in Fig. 5C, red open triangle in Fig. 5D). The spine is present again when re-imaged at 14 mW, confirming that its absence in the 10 mW image was not due to bleaching (Fig. 5E). For the PSTPM images used in the current study, laser power was adjusted to maintain photon counts within the linear range of the PMT along the shaft of the brightest dendrites in the field of view (see Discussion), and the results shown in Fig. 5 suggest that the dimmest spines are not always visible using these imaging parameters. Thus, we conclude that the higher laser power used for TFM imaging, when combined with the reconstruction capabilities of the trained network, can reveal dim spines not visible with the low excitation power used for PSTPM.

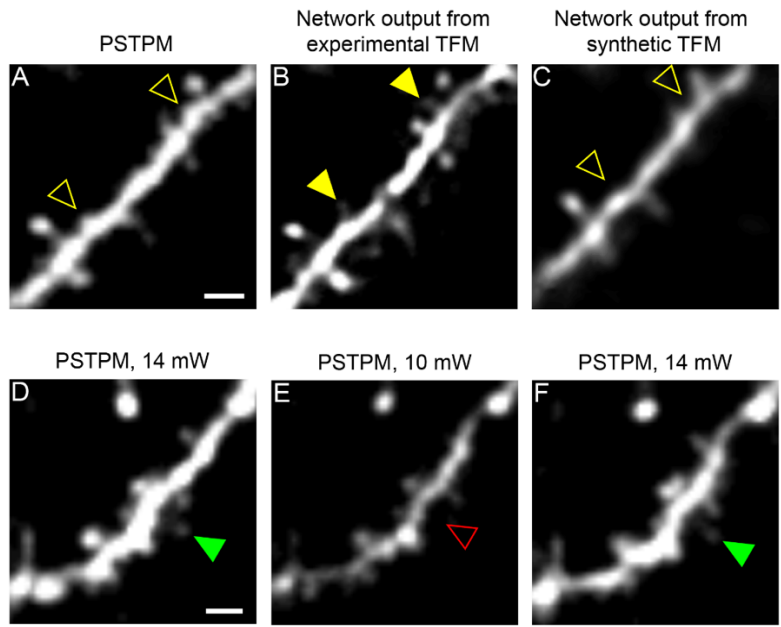

Fig. 5 False positive analysis, including numerical (A-C) and experimental (D-F) tests. A-C) For the same branch in Fig. 3C, synthetic TFM images are generated from the PSTPM images and used as input of the trained network. The false-positive spines (marked by filled yellow triangles in B, yellow circles in Fig. 3C) are missing in the output when using synthetic TFM stack as inputs (C). D-F) Dim spines can be missed in PSTPM images depending on excitation power. The same segment of dendrite was imaged using 14 mW laser power (C), 10 mW laser power (D), and then 14 mW again (E). Arrows denote a dim spine that would be marked as present in the 14 mW images (green filled arrows) but absent in the 10 mW image (red open arrow). The second 14 mW image was taken after the 10 mW image to confirm that failure to see the dim spine was not due to bleaching. Scale bars, 2 $\mu m$.

We also used the NN to process the TFM images of the same dendritic branch experimentally imagined (with TFM), and synthetically generated from PSTPM images (using the forward model). As presented in Fig. 5, "false-positive" spines are present only in the reconstruction from experimental TFM inputs (Fig. 5B); no false positives are seen in the reconstruction from the synthetic TFM inputs (Fig. 5C). In summary, the seemingly false-positive spines in the processed TFM measurements represent actual small spines seen with TFM but not PSTPM, rather than artifacts generated by the trained network.

**Discussion and Summary**

Our results indicate that the presented 3D neural network model can enable fast imaging of large 3D volumes at high resolution, recovering small, biologically relevant structures that would be lost to scattering in traditional TFM. Using both simulation and experimental data, we show that our trained neural network can reconstruct dendritic spines from TFM images of layer 2/3 pyramidal neurons in the mouse visual cortex. Using PSTPM of the same dendrites as ground truth to test the performance of our trained network, we determine that our trained network recovers 91.2% of the spines present in the PSTPM data, while only 49.8% of those spines are detectable in the TFM data. Further quantification of fluorescence intensity shows that the trained network produces clear peaks in fluorescence at the sites of dendritic spines, similar to those present in the PSTPM image, while those same spines are often buried in the noise of scattered fluorescence in the TFM image.

Dendritic spines house the majority of excitatory synapses[41], and their dynamics can represent critical events in circuit reorganization across development[42-45] and in response to specific experiences[46-49]. Historically, imaging of spine dynamics has been performed in anesthetized animals using relatively long imaging sessions due to the time-consuming nature of performing PSTPM across large volumes at sufficient resolution to capture spines, which can be less than 1 micron in length[5,42-49]. The PSTPM image shown in Fig. 3J represents a volume of 2,560,000 $\mu m^3$ and took 30 minutes to acquire. The long duration of PSTPM imaging sessions limits these studies to capturing plasticity events that occur between sessions, when animals are awake, often over relatively long timescales. Visualization of spine and synapse dynamic events occurring on a faster time scale is not possible with such slow imaging speeds.

While faster imaging approaches such as traditional TFM have been used for imaging large, bright structures such as neuronal cell bodies[10,11], they have not been used for imaging subcellular structures because the fluorescence from small, dim structures is lost to scattering. Indeed, in Fig. 3K, we show that in the same 2,560,000 $\mu m^3$ tissue volume imaged in less than 1 minute using TFM, more than half of the spines are undetectable. However, with our trained 3D convolution neural network, we were able to recover more than 90% of spines, producing a high-resolution output image that matched closely with the ground truth PSTPM data. As an exact comparison of imaging time, to obtain a stack with dimensions of $800 \times 800 \times 64$ pixels, it takes approximately 27.3 minutes for PSTPM, but only 32 seconds for TFM.

Thus, we demonstrate that using a deep learning approach combined with TFM, we can image dendritic spines across a large tissue volume orders of magnitude faster than with PSTPM. Our approach produces high-quality images without the need for long imaging sessions. Fast, high-resolution imaging of large tissue volumes opens the door to new experimental paradigms in which structural plasticity events can be assayed over shorter timescales and potentially in awake, behaving animals.

**Methods**
**3D Network Architecture** In this work, a 3D convolutional neural network is used to implement the proposed learning method (Supplementary Figure 1). Similar to the approach presented in[50,51], in the left path of the neural network, 3D convolutions are performed at each stage to extract the features from the data, and at the end of each stage 3D Max pooling is used for downsampling purposes. In the right path of the neural network, 3D convolutions are also used to extract the features at each stage, whereas at the end of each stage, transposed 3D convolutions are used to improve the resolution. Both the left and right paths of the network consist of four stages with a connecting middle stage. Each stage includes two to three 3D convolutions coupled with rectified linear unit (ReLU) activation function and batch normalization. Skip connections with concatenations are used between the left and right path to keep the features contained in the left path. Since the inputs and outputs share many important features, a long skip connection is further added to connect the input all the way to the output, forming a residual learning[52].

In this work, the input data have sizes of $128 \times 128 \times 64$ voxels, and the convolutions performed in the network use volumetric kernels with $3 \times 3 \times 3$ voxels. In the left path, by performing 3D Max pooling at the end of each stage with a pool size of (2, 2, 2), the resulting resolution of feature maps are halved when data proceed through. Meanwhile, due to the number of channels doubled by convolutions at the beginning of each stage, the number of feature maps are doubled. In the right path, the Max pooling is replaced by the transpose 3D convolutions with strides of (2, 2, 2). Thus, when data pass through, the spatial resolution of the feature maps is doubled, and information is gathered to finally output a high-resolution cube with $128 \times 128 \times 64$ voxels.

**Generating Training and Testing Data** In the training process, a total number of 33 PSTPM image stacks provided ground truth for the training data, which are concatenated as 3D stacks from *in vivo* images acquired from the brains of 26 mice at different depths. Due to inhomogeneity of the brain tissue that also includes blood vessels and meninges, the measured samples are heterogeneous. Each PSTPM image stack is interpolated into a dimension of $922 \times 922 \times 64$ pixels. When generating training data, 134 small stacks with a dimension of $128 \times 128 \times 64$ pixels are extracted from each PSTPM image stack, where the central positions of these 134 small stacks are randomly distributed in the original PSTPM image stacks, as presented in Fig. 1. Consequently, there are a total number of $134 \times 33 = 4422$ stacks used as ground truth in the training process, of which 95% are used as training data and 5% are used as validation data.

In the training process, the inputs of the neural network, i.e., the TFM image stacks $I_T$, are synthetically generated from the PSTPM image stacks $I_P$. Specifically, we have $I_T(z) = P[I_P(z) * P_s(z)]$. Here, the operators $P$ and $*$ represents adding Poisson noise and convolution operations, respectively. $P_s(z)$ is the scattering point spread function (sPSF) at depth $z$ and it is simulated according to the method proposed by Kim et al.[2], where the sPSF consists of two components including the ballistic photon distribution and the scattered photon distribution (Supplementary Section 2).

In the testing process, both synthetic data and experimental data are used. Firstly, the synthetic data are generated in the same way as those in the training data, where the input stack is synthetically generated from the experimental PSTPM stack. It is noted that the PSTPM stack used in the test has not been used in the training process. Secondly, experimental data are used, where the input and output stacks are measured *in vivo* from TFM and PSTPM, respectively. Due to the limited memory of our hardware (Dell with Xeon Silver 4114 CPU, 32GB RAM, NVIDIA TITAN RTX GPU 24 GB), both the inputs and outputs have dimensions of $128 \times 128 \times 64$. Nevertheless, due to the fast speed of the reconstruction, we can reconstruct large dimensional stacks by sticking multiple outputs into a single large stack.

**Neural Network Parameters** Mean squared error is used as loss function in this work. The hyperparameters for the network are as follows: The learning rate is $10^{-4}$. Adam optimizer is used for optimization. Maximum 30 epochs with 300 steps for each epoch are set for training. The batch size is 3 and channel size is 1. To mitigate the effects of possible overfitting, we empirically apply an "early stopping" strategy. Specifically, the training process is stopped when there is no obvious change in the loss of validation data (Supplementary Figure 2).

**Preparation of *In Vivo* Specimens** All animal procedures were approved by the Massachusetts Institute of Technology Committee on Animal Care and meet the NIH guidelines for the care and use of vertebrate animals. To enable visualization of individual neurons' dendritic morphology, *in utero* electroporations were performed on embryonic day 15.5 timed pregnant C57BL/6J mice to express a fluorescent protein in a sparse population of layer 2/3 pyramidal neurons. Constructs used for *in utero* electroporation were a Cre-dependent mScarlet-I cell fill or a Cre-dependent eYFP cell fill[5] at a concentration of 0.7 $\mu g/\mu l$, along with a Cre plasmid[53] at a concentration of 0.03 $\mu g/\mu l$. Fast Green (0.1%) was included in the plasmid solution for visualization. A total of 0.5-1.0 $\mu l$ of the plasmid solution was injected into the right lateral ventricle with a 32 G Hamilton Syringe (Hamilton Company), and five pulses of 36 V (duration 50 ms, frequency 1 Hz) targeting the visual cortex were delivered from a square-wave electroporator (ECM830, Harvard Apparatus). Pups were implanted with a 5 mm cranial window over the right hemisphere as described previously[54] and fitted with a custom head mount to enable fixation to the microscope stage. All imaging took place under isoflurane anesthesia (1.25%) with the head mount fixed to the microscope stage. The animal whose data are shown in Fig. 3 expressed mScarlet-I as a cell fill, underwent cranial window surgery at postnatal day 14, and was imaged at postnatal day 32.

**Collection of PSTPM Images** *In vivo* PSTPM was performed on a custom-built microscope to visualize the dendritic morphology of layer 2/3 pyramidal neurons in the visual cortex of anesthetized mice. The source of excitation was a Mai Tai HP Ti: Sapphire laser (Rep rate 80 MHz, Spectra Physics) tuned to 1030 nm. The average power delivered to the specimen ranged from 10 to 75 mW depending on cell brightness and imaging depth. Galvanometric XY scanning mirrors (6215H, Cambridge Technology) and a piezo actuator Z positioning system (Piezosystem Jena) were used for XY and Z movement, respectively. The dwell time per pixel was 40 $\mu s$. The pixel size was 0.25 $\mu m$ in XY, and the Z step size was 1 $\mu m$. The beam was focused by a 20x/1.0 NA water immersion objective lens (W Plan-Apochromat, Zeiss). Emissions were collected by the same objective lens and passed through an IR blocking filter (E700SP, Chroma Technology). Emissions were then separated by dichroic mirrors at 520 nm and 560 nm. After passing through three independent bandpass filters (485/70 nm, 550/100 nm, and 605/75 nm), emissions were collected simultaneously onto three separate PMTs. Raw 2-photon scanning data were processed for spectral linear unmixing and converted into a tif Z stack using Matlab (Mathworks) and ImageJ (NIH). For mice expressing an eYFP cell fill, the 550/100 nm (i.e. yellow) channel was used to visualize the dendritic morphology. For mice expressing an mScarlet-I cell fill, the 605/75 nm (i.e. red) channel was used to visualize the dendritic morphology.

**Collection of TFM Images** *In vivo* TFM was performed on a custom-built microscope using a 1035 nm fixed wavelength laser (repetition rate 1 MHz, spectrum width $\pm 5$ nm, Monaco, Coherent) as the source of excitation. The average power delivered to the specimen was 620 mW. The beam was mechanically scanned along the y-axis by a scanning mirror (6350, Cambridge Technology, MA, USA). A cylindrical lens (f=150 mm) focused the beam into a line on a grating (20RG1200-1000-2, Newport Co., CA, USA, 1200 grooves/mm), which generated dispersion along the x-axis. A diffractive optical element (DOE) split the beam into 4 beams in the Fourier plane. After the tube lens and objective lens (XLUMPlanFL, 20x, 0.95NA, Olympus), the diffracted beam focused 4 scanning lines on the imaging plane. The image was detected by 2 sCMOS cameras simultaneously (red channel, Prime95B, Photometrics; yellow channel, PCO edge 5.5, PCO AG). Emissions for the red and yellow channels were separated by a dichroic mirror at 560 nm. Emissions collected in the red channel were used to visualize dendritic morphology labeled by the mScarlet-I cell fill. The field of view was $250 \times 250$ $\mu m^2$, the z step size was 1 $\mu m$, the pixel size was 170 nm, and the exposure time was 500 ms per plane.

**Acknowledgements**

This work was supported by the Center for Advanced Imaging at Harvard University (DNW, ZW), 5-P41EB015871-32 (ZW), R21 NS105070 (EN, PS), and F32 MH115441 (JB). DNW is also supported by the John Harvard Distinguished Science Fellowship Program within the FAS Division of Science of Harvard University.

**Authors' contribution**

Z. W. and D. N. W. conceived the idea. J. R. B. and Y. X. designed and conducted the experiments. Z. W. and J. R. B. analyzed the data and wrote the manuscript. X. C, P. T. C. So, and E. N. participated in the discussions and contributed with valuable suggestions. D. N. W. supervised the project. All authors read and corrected the manuscript before manuscript submission.

**Additional information**

Supplementary information is available in the online version of the paper. Correspondence and requests for data and code should be addressed to Dushan N. Wadduwage (Email: wadduwage@fas.harvard.edu).

**Competing financial interests:** The authors declare no competing financial interests.

**Data availability.** The data that support the plots within this paper and other findings of this study are available from the corresponding authors upon reasonable request.

**Code availability.** The codes that support the plots within this paper and other findings of this study are available from the corresponding authors upon reasonable request.